\documentclass{bmcart}

\usepackage{amsthm,amsmath,graphicx}
\usepackage{xcolor} %color
\usepackage{comment}
\usepackage{multirow}
\usepackage{todonotes}
\usepackage{enumitem}% abbrv

\startlocaldefs
\newlist{abbrv}{itemize}{1}
\setlist[abbrv,1]{label=,labelwidth=1in,align=parleft,itemsep=0.1\baselineskip,leftmargin=!}
\endlocaldefs

\begin{document}

%%% Start of article front matter
\begin{frontmatter}

\begin{fmbox}
\dochead{Research}


\title{Automated quantification of myocardial tissue characteristics from native T\textsubscript{1} mapping using  neural networks with Bayesian inference for uncertainty-based quality-control}

%%%%%%%%%%%%%%%%%%%%%%%%%%%%%%%%%%%%%%%%%%%%%%
%%                                          %%
%% Enter the authors here                   %%
%%                                          %%
%% Specify information, if available,       %%
%% in the form:                             %%
%%   <key>={<id1>,<id2>}                    %%
%%   <key>=                                 %%
%% Comment or delete the keys which are     %%
%% not used. Repeat \author command as much %%
%% as required.                             %%
%%                                          %%
%%%%%%%%%%%%%%%%%%%%%%%%%%%%%%%%%%%%%%%%%%%%%%
\author[
   addressref={aff1},                   % id's of addresses, e.g. {aff1,aff2}
   corref={aff1},                       % id of corresponding address, if any
   %noteref={n1},                        % id's of article notes, if any
   email={esther.puyol\textunderscore anton@kcl.ac.uk}   % email address
]{\inits{EPA}\fnm{Esther} \snm{Puyol-Ant{\'o}n}}
\author[
   addressref={aff1,aff2},
   email={jacobus.ruijsink@kcl.ac.uk}
]{\inits{BR}\fnm{Bram} \snm{Ruijsink}}
\author[
   addressref={aff3},
   email={baumgartner@vision.ee.ethz.ch}
]{\inits{CFB}\fnm{Christian F.} \snm{Baumgartner}}%
\author[
   addressref={aff4},
   email={m.sinclair@imperial.ac.uk}
]{\inits{MS}\fnm{Matthew} \snm{Sinclair}}
\author[
   addressref={aff3},
   email={ender.konukoglu@vision.ee.ethz.ch}
]{\inits{EK}\fnm{Ender} \snm{Konukoglu}}
\author[
   addressref={aff1,aff2},
   email={reza.razavi@kcl.ac.uk}
]{\inits{RR}\fnm{Reza} \snm{Razavi}}
\author[
   addressref={aff1},
   email={andrew.king@kcl.ac.uk}
]{\inits{APK}\fnm{Andrew P.} \snm{King}}

%%%%%%%%%%%%%%%%%%%%%%%%%%%%%%%%%%%%%%%%%%%%%%
%%                                          %%
%% Enter the authors' addresses here        %%
%%                                          %%
%% Repeat \address commands as much as      %%
%% required.                                %%
%%                                          %%
%%%%%%%%%%%%%%%%%%%%%%%%%%%%%%%%%%%%%%%%%%%%%%
\address[id=aff1]{%                           % unique id
  \orgname{School of Biomedical Engineering \& Imaging Sciences, King\rq{}s College London}, % university, etc
  \street{Rayne Institute, 4th Floor Lambeth Wing St Thomas Hospital, Westminster Bridge Road}, %                
  \postcode{SE1 7EH}                        % post or zip code
  \city{London},                              % city
  \cny{UK}                                    % country
}
\address[id=aff2]{%
  \orgname{Department of Adult and Paediatric Cardiology, Guy\rq{}s and St Thomas\rq{} NHS Foundation Trust},
  \city{London},                              % city
  \cny{UK}  
}
\address[id=aff3]{%
  \orgname {Computer Vision Lab, ETH Z{\"u}rich},
  \street{Sternwartstrasse 7},
  \postcode{Z{\"u}rich}
  \city{London},                              % city
  \cny{Switzerland}  
}
\address[id=aff4]{%
  \orgname {Biomedical Image Analysis Group, Department of Computing, Imperial College London},
  \street{3rd floor Huxley Building, 180 Queen\rq{}s Gate},
  \postcode{SW7 2AZ}
  \city{London},                              % city
  \cny{UK}  
}

%%%%%%%%%%%%%%%%%%%%%%%%%%%%%%%%%%%%%%%%%%%%%%
%%                                          %%
%% Enter short notes here                   %%
%%                                          %%
%% Short notes will be after addresses      %%
%% on first page.                           %%
%%                                          %%
%%%%%%%%%%%%%%%%%%%%%%%%%%%%%%%%%%%%%%%%%%%%%%

%\note{Sample of title note}     % note to the article
%\note[id=n1]{Equal contributor} % note, connected to author

\end{fmbox}% comment this for two column layout

%%%%%%%%%%%%%%%%%%%%%%%%%%%%%%%%%%%%%%%%%%%%%%
%%                                          %%
%% The Abstract begins here                 %%
%%                                          %%
%% Please refer to the Instructions for     %%
%% authors on http://www.biomedcentral.com  %%
%% and include the section headings         %%
%% accordingly for your article type.       %%
%%                                          %%
%% The Abstract should not exceed 350 words. Please minimize the use of abbreviations and do not cite references in the abstract. Reports of randomized controlled trials should follow the CONSORT extension for abstracts. The abstract must include the following separate sections:

%%Background: the context and purpose of the study
%%Methods: how the study was performed and statistical tests used
%%Results: the main findings
%%Conclusions: brief summary and potential implications
%%%%%%%%%%%%%%%%%%%%%%%%%%%%%%%%%%%%%%%%%%%%%%

\begin{abstractbox}

\begin{abstract} % abstract
\parttitle{Background} 
Tissue characterisation with cardiovascular magnetic resonance (CMR) parametric mapping has the potential to detect and quantify both focal and diffuse alterations in myocardial structure not assessable by late gadolinium enhancement. Native T\textsubscript{1} mapping in particular has shown promise as a useful biomarker to support diagnostic, therapeutic and prognostic decision-making in ischaemic and non-ischaemic cardiomyopathies. 

\parttitle{Methods}
Convolutional neural networks (CNNs) with Bayesian inference are a category of artificial neural networks which model the uncertainty of the network output. This study presents an automated framework for tissue characterisation from native ShMOLLI  T\textsubscript{1} mapping at 1.5 Tesla using a Probabilistic Hierarchical Segmentation (PHiSeg) network \cite{baumgartner2019}. In addition, we use the uncertainty information provided by the PHiSeg network in a novel automated quality control (QC) step to identify uncertain T\textsubscript{1} values.  The PHiSeg network and QC were validated against manual analysis on a cohort of the UK Biobank containing healthy subjects and chronic cardiomyopathy patients (N=100 for the  PHiSeg network and N=700 for the QC). We used the proposed method to obtain reference T\textsubscript{1} ranges for the left ventricular myocardium in healthy subjects as well as common clinical cardiac conditions.

\parttitle{Results}
T\textsubscript{1} values computed from automatic and manual segmentations were highly correlated (r=0.97). Bland-Altman analysis showed good agreement between the automated and manual measurements. The average Dice metric was 0.84 for the left ventricular myocardium. The sensitivity of detection of erroneous outputs was 91\%. Finally, T\textsubscript{1} values were automatically derived from 14,683 CMR exams from the UK Biobank.

\parttitle{Conclusions} 
The proposed pipeline allows for automatic analysis of myocardial native T\textsubscript{1} mapping and includes a QC process to detect potentially erroneous results.  T\textsubscript{1} reference values were presented for healthy subjects and common clinical cardiac conditions from the largest cohort to date using T\textsubscript{1}-mapping images.

\end{abstract}

%%%%%%%%%%%%%%%%%%%%%%%%%%%%%%%%%%%%%%%%%%%%%%
%%                                          %%
%% The keywords begin here                  %%
%%                                          %%
%% Put each keyword in separate \kwd{}.     %%
%%                                          %%
%%%%%%%%%%%%%%%%%%%%%%%%%%%%%%%%%%%%%%%%%%%%%%

\begin{keyword}
\kwd{Native T\textsubscript{1} Mapping}
\kwd{Convolutional Neural Networks}
\kwd{Automatic Analysis}
\kwd{Cardiac MR Segmentation}
\kwd{Quality Control}
\kwd{UK Biobank}
\end{keyword}

\end{abstractbox}

\end{frontmatter}

%%%%%%%%%%%%%%%%%%%%%%%%%%%%%%%%%%%%%%%%%%%%%%
%%                                          %%
%% The Main Body begins here                %%
%%                                          %%
%% Please refer to the instructions for     %%
%% authors on:                              %%
%% http://www.biomedcentral.com/info/authors%%
%% and include the section headings         %%
%% accordingly for your article type.       %%
%%                                          %%
%% See the Results and Discussion section   %%
%% for details on how to create sub-sections%%
%%                                          %%
%% use \cite{...} to cite references        %%
%%  \cite{koon} and                         %%
%%  \cite{oreg,khar,zvai,xjon,schn,pond}    %%
%%  \nocite{smith,marg,hunn,advi,koha,mouse}%%
%%                                          %%
%%%%%%%%%%%%%%%%%%%%%%%%%%%%%%%%%%%%%%%%%%%%%%

%%%%%%%%%%%%%%%%%%%%%%%%% start of article main body
% <put your article body there>

%%%%%%%%%%%%%%%%
%% Background %%
%%%%%%%%%%%%%%%%

\section*{Background/Introduction}
% Clinical Motivation
Cardiovascular magnetic resonance (CMR) provides insights into myocardial structure and function noninvasively, with high diagnostic accuracy and without ionising radiation. Late Gadolinium Enhancement (LGE) has become the reference standard for non-invasive imaging of myocardial scar and focal fibrosis in both ischaemic \cite{kim2000use} and non-ischaemic cardiomyopathy \cite{patel2017role}. LGE is useful in cardiac conditions which have stark regional differences within the myocardium, but it cannot correctly visualise myocardial pathology that is diffuse in nature and affects the myocardium uniformly. Examples include diffuse myocardial inflammation, fibrosis, hypertrophy, and infiltration \cite{sado2013identification}. In contrast, native T\textsubscript{1}-mapping provides quantitative myocardial tissue characterisation, without the need for gadolinium \cite{moon2013myocardial}. Previous work has shown that T\textsubscript{1} mapping can help to detect diffuse myocardial disease in early disease stages and aids in diagnosing the diseases' underlying cardiac dysfunction. \\

% T1 mapping correction
Despite its recognised potential, T\textsubscript{1} mapping analysis typically requires time-consuming manual segmentation of T\textsubscript{1} maps. Moreover, external factors, such as hematocrit and blood flow, impact the obtained values and create variability that reduces the ability to separate healthy from diseased myocardium. Several blood correction models have been proposed to limit the impact of external factors \cite{nickander2017blood,shang2018blood,reiter2013normal}. However, these methods have not been evaluated in large cohort studies. Automating T\textsubscript{1} analysis of myocardial tissue characterisation sequences could facilitate the clinical use of T\textsubscript{1} mapping and unlock the potential to obtain T\textsubscript{1} data in large populations.\\

% Solution to the problem and impact
In recent years, deep learning methods have shown great success in segmenting anatomical and pathological structures in medical images \cite{bai2018automated, ruijsink2019fully, fahmy2019automated}. For many tasks, their accuracy is comparable to human-level performance, or even surpasses it. In the context of CMR imaging,  semi-automatic and automatic techniques for cardiac cine \cite{ruijsink2019fully, bai2018automated} and flow \cite{goel2014fully} imaging have been developed. One paper has proposed an automated segmentation method for native T\textsubscript{1} maps \cite{fahmy2019automated}. However, this method only extracted global left ventricle (LV) myocardial T\textsubscript{1} values, whereas regional assessment of septal and/or focal lesion T\textsubscript{1} values is typically used to characterise diseases \cite{liu2017measurement, messroghli2017clinical}. Furthermore, T\textsubscript{1} values were only reported for healthy subjects and a pooled group of cardiovascular diseases (CVD), without distinguishing between different myocardial disease processes and these values were not corrected for myocardial blood volume. In this paper we add further insight into the aforementioned areas.\\

Medical segmentation problems are often characterised by ambiguities, some of them inherent to the data such as poor contrast, inhomogeneous appearance and variations in imaging protocol, and some due to inter- and intra-observer variability in the annotated data used for training.  To limit the effect of these factors and detect failed cases, some groups have proposed to incorporate quality control (QC) techniques \cite{fahmy2019automated, ruijsink2019fully, robinson2019automated}. We believe that modeling uncertainty at a per-pixel level is an important step in understanding the reliability of the segmentations and increasing clinicians' trust in the model's outputs. Several works have investigated uncertainty estimation for deep neural networks \cite{kendall2017uncertainties, lakshminarayanan2017simple, zhu2018bayesian}. A popular approach  to account for the uncertainty in the learned model parameters is to use variational Bayesian methods, which are a family of techniques for approximating Bayesian inference over the network weights. These methods can be used to automatically segment the anatomy of interest, but additionally provide a pixel-wise uncertainty map of the confidence of the model in segmenting the input image. Budd \textit{et al} \cite{budd2019confident} proposed to use this approach to automatically estimate fetal Head Circumference from Ultrasound imaging and provide real-time feedback on measurement robustness.\\

% Solution
In this paper, we develop a tool for automated segmentation and analysis of T\textsubscript{1} maps. We use the PHiSeg network \cite{baumgartner2019} to segment the images, and additionally use the generated uncertainty information in a novel QC process to identify uncertain (and potentially inaccurate) segmentations. To the best of our knowledge this is the first time that segmentation uncertainty information has been used for QC in medical image segmentation. 
By incorporating this QC process, our framework automatically controls the quality of the segmentations and rejects those that are uncertain. We hypothesise that this method can be used to derive high quality T\textsubscript{1} data without human interaction from large-scale databases. Using the proposed method we compute mean global and regional native T\textsubscript{1} values from 14,683 subjects from the UK Biobank, which represents the largest cohort for T\textsubscript{1} mapping images to date. We report reference values for healthy subjects and interrogate typical values obtained in important relevant subgroups of cardiomyopathies. In addition, we investigate if a blood correction model for T\textsubscript{1} \cite{nickander2017blood} provides better discrimination between healthy and diseased myocardium. \\

\section*{Materials and Methods}

\subsection*{UK Biobank dataset}
CMR imaging was carried out on a 1.5 Tesla scanner (Siemens Healthcare, Erlangen, Germany). For each subject, the Shortened Modified Look-Locker Inversion recovery technique (ShMOLLI, WIP780B) was used to perform native (non-contrast) myocardial T\textsubscript{1} mapping in a single mid-ventricular short axis (SAX) slice (TE/TR/flip-angle (FA): 1.04ms/ 2.6ms/ 35$^{\circ}$, voxel size 0.9 x 0.9 x 8.0 mm).  The matrix size of all images was unified to 192 x 192. Details of the full image acquisition protocol can be found in \cite{petersen2015uk}. \\

As pre-processing, all of the CMR DICOM images were converted into NIfTI format. Training and validation data were obtained through manual segmentation of the images by an experienced CMR cardiologist. The LV endocardial and epicardial borders and the right ventricle (RV) endocardial border were  traced using the ITK-SNAP interactive image visualisation and segmentation tool \cite{yushkevich2006user}.\\

From the UK Biobank database, we first excluded any subject with systemic disease. Subsequently, we identified patients with CVD from the included cohort using ICD10 codes. We included 11 relevant groups of CVD: acute myocarditis, aortic stenosis (AS), atrial fibrillation (AF), cardiac sarcoidosis, chronic coronary artery disease (CAD), dilated cardiomyopathy (DCM), hypertrophic cardiomyopathy (HCM), pheochromocytoma, obsesity or takotsubo cardiomyopathy. Of the remaining subjects, those who had no history of CVD nor any cardiovascular risk factors were included as healthy subjects. \\

From the selected study population, we randomly chose three sub-cohorts: a set of 800 subjects (consisting of both healthy subjects as well as subjects with a wide variety of CVDs) for training, a set of 100 subjects (50 healthy subjects and 50 chronic cardiomyopathy subjects) for validation of the segmentations provided by the PHiSeg network, and a set of 700 subjects (500 healthy subjects and 200 chronic cardiomyopathy subjects) for the validation of the QC process.

\subsection*{Automated image analysis}
The proposed workflow for automated T\textsubscript{1} map analysis is summarised in Figure \ref{fig:Fig1_pipeline} and described in detail in the following subsections. 

\subsubsection*{Deep neural network with Bayesian inference for segmentation}
In this work, we used a Probabilistic Hierarchical Segmentation (PHiSeg) network \cite{baumgartner2019}, a recently proposed deep learning network with Bayesian inference for segmentation of the LV blood pool, LV myocardium and RV blood pool from T\textsubscript{1} mapping images (Figure \ref{fig:Fig1_pipeline}). The PHiSeg network employs convolutions to learn task-specific representations of the input data and predicts a pixel-wise segmentation from an input image based on this representation. In addition, an uncertainty map is generated which quantifies the pixel-wise uncertainty of the segmentation.\\

The PHiSeg network \cite{baumgartner2019} models the segmentation problem at multiple scales from fine to coarse. Performing inference with this model using a conditional variational autoencoder approach results in a network architecture resembling the commonly used U-Net. However, in contrast to a U-Net, this network allows modelling of the joint probability of all pixels in the segmentation map. Specifically, it allows sampling multiple plausible segmentation hypotheses for an input image. In this manner, in addition to producing a per pixel prediction of the label class, it also allows estimation of the uncertainty corresponding to each pixel. The network architecture features a number of convolutional layers, each using a 3x3 kernel and  a rectified linear unit (ReLU) activation function. After every three convolutions, the feature map is downsampled by a factor of 2 to learn more global scale features. After performing probabilistic inference at each level, the learned features are upsampled and fused to produce a predicted segmentation mask and a uncertainty map at the original image resolution. To train the model, we aim to find the neural network parameters which maximise the evidence lower bound (ELBO), which models the marginal likelihood of the observed data. The general idea is that a higher marginal likelihood for a given model indicates a better fit of the data by that model and hence a greater probability that the model in question was the one that generated the data \cite{yang2017understanding}. A detailed description of the method, as well as the network architecture can be found in \cite{baumgartner2019}.\\

\subsubsection*{Network training and testing}
For training of the network, all images were cropped using the manual segmentation to the same size of 192$\times$192 and intensity normalised to the range of [0,1]. Data augmentation was performed on-the-fly using random translations, rotations, scalings and intensity transformations to each mini-batch of images before feeding them to the network. Each mini-batch consisted of 20 native T\textsubscript{1} images. To optimise the loss function we used the Adam optimiser, with the momentum  set to 0.9 and the learning rate to $10^{-3}$. The models were trained for 50,000 iterations on a NVIDIA GeForce GTX TITAN GPU and the model with highest average Dice score (on the validation set) over all classes was selected.

\subsubsection*{Generating uncertainty maps}
During test time, we used the PHiSeg network to sample $T$ different segmentation output samples for a single given input (we used $T$=100). From these multiple segmentations the final predicted segmentation was calculated as the average softmax probability over all of the segmentation samples, and the uncertainty map was generated by computing the cross entropy between the mean segmentation mask and the segmentation samples.

\subsubsection*{Quality control}
Our QC process comprises two steps, both based upon different aspects of the uncertainty information provided by the PHiSeg network. To train these QC steps, we manually labelled the PHiSeg-obtained segmentations as correct or incorrect in a cohort of 800 subjects (consisting of both healthy subjects as well as subjects with a wide variety of CVDs).\\

First, we used the ELBO output \cite{baumgartner2019, yang2017understanding} of the trained PHiSeg network to reject uncertain segmentations. The ELBO quantifies how likely it is that the segmentation is correct. We used the manual labellings to determine a threshold and any ELBO value above this threshold resulted in the segmentation being rejected.\\

The second QC step is defined as an image classification problem, where each image/segmentation pair is classified as accurate or inaccurate. The outputs of the PHiSeg network (i.e. the segmentation and uncertainty map) were used as input to a deep learning image classifier. For the image classifier, we used a VGG-16 CNN network \cite{simonyan2014very}, which consists of a stack of convolutional layers followed by three fully-connected layers for classification. Each convolutional layer uses a 3x3 kernel and is followed by batch normalisation and ReLU. Details of the VGG-16 network can be found in \cite{simonyan2014very}.\\

Data are rejected if either of these two QC steps fails. The combination of these two steps ensures that T\textsubscript{1} images acquired on different planes or with inaccurate segmentations are identified and rejected for further analysis.

\subsubsection*{T\textsubscript{1} map analysis}
Myocardial T\textsubscript{1} values were measured from the mid-ventricular SAX slice for the whole myocardium, as well as for the interventricular septum and free-wall segments separately. From the predicted segmentation, the RV-LV intersection points were automatically detected from the LV/RV segmentation masks (RV1 and RV2 in Figure \ref{fig:Fig1_pipeline}) using the hit-or-miss transform, which is a morphological operation that detects a given configuration (or pattern) in an image. In our case, we aimed to detect the intersection between background, LV myocardium and RV labels. These RV-LV intersections were used to divide the LV myocardium mask into a LV interventricular septum (LVIVS) mask and a LV free-wall (LVFW) mask.

\subsubsection*{Myocardial T\textsubscript{1} blood correlation}
For myocardial T\textsubscript{1} blood correlation, we used the model proposed by Nickander \textit{et al} \cite{nickander2017blood}, which used a linear correlation between myocardial T\textsubscript{1} and blood measurements as follows:
\begin{equation}
    \text{T\textsubscript{1}\textsuperscript{corrected}} = \text{T\textsubscript{1}\textsuperscript{uncorrected}} + \alpha \cdot(\text{R\textsubscript{mean}} - \text{R\textsubscript{patient}}),
    \label{eq:blood_correction}
\end{equation}
where T\textsubscript{1}\textsuperscript{uncorrected} is the native myocardial T\textsubscript{1} value, R\textsubscript{mean} is the mean R\textsubscript{1} for the patient cohort, and $\alpha$ is calculated as the slope of the linear regression between myocardial T\textsubscript{1} and blood T\textsubscript{1} measurements.\\

The blood T\textsubscript{1} value was computed from RV and LV blood pool regions of interest (ROI) in the mid-ventricular SAX T\textsubscript{1} map. To generate the LV and RV ROIs we eroded the LV/RV blood pool segmentations to generate a mask that has 1/3 of the area of the original mask (see Figure \ref{fig:Fig2_ROI_T1_corr}). To ensure that no papillary muscles or trabeculae were included, we rejected any pixel whose T\textsubscript{1} value was less than 1.5 times the interquartile range below the first quartile of the blood pool values. The blood T\textsubscript{1} value was calculated as the mean of the LV and RV values calculated in this way, and then converted to the T\textsubscript{1} relaxation rate (R\textsubscript{1}=1/T\textsubscript{1}). \\

\subsection*{Reference values}
In total, we analysed CMR scans of 14,683 subjects (62 $\pm$ 22 yrs., 48\% males) included in the UK Biobank cohort using our method. First, we derived reference values in 5,685 healthy subjects, selected using stringent criteria to exclude any disease or risk factor that impacts the heart or vasculature (see details in our previous paper \cite{ruijsink2019fully}). Next, we analysed data of patients known to have one of 11 different CVD's. We also obtained indexed LV end diastolic volume (iLVEDV), LV ejection fraction (LVEF) and indexed LV mass (iLVM) from cine CMR data, using the automated method described in \cite{ruijsink2019fully}. Outliers were defined a priori as values 3 interquartile ranges below the first or above the third quartile and they were removed from the analysis.

\subsubsection*{Evaluation of the method}
We evaluated the performance of the automated network as follows:\\

\textbf{Deep neural network with Bayesian inference}: To validate the PHiSeg network, a cohort of 50 healthy volunteers and 50 chronic cardiomyopathy patients was selected and manually segmented. These subjects were not used for training the PHiSeg network. We used the Dice metric to measure the degree of overlap between the automated and manual segmentations. The Dice metric has values between 0 and 1, where 0 denotes no overlap and 1 denotes perfect agreement.
Furthermore, Bland-Altman analysis and Pearson’s correlation were used to compare the obtained global LV native myocardial T\textsubscript{1} values, and the  T\textsubscript{1} values in the LVIVS and LVFW, between the automated and manual segmentations. \\

\textbf{Quality control}: To assess the accuracy of the QC process, we manually labelled the PHiSeg obtained segmentations as correct or incorrect in a cohort of 500 healthy volunteers and 200 chronic cardiomyopathy patients, which are independent from the training cohort. We computed sensitivity (\% of manually labelled as incorrect image/segmentation pairs that were correctly detected), specificity (\% of manually labelled as correct image/segmentation pairs that were correctly identified), and balanced accuracy (averaged percentages of correct answers for correct/incorrect classes individually).

\subsubsection*{Statistical analysis}
Statistical analysis was performed using Statsmodels, a Python library for statistical and econometric analysis \cite{seabold2010statsmodels}. Normality of distributions was tested with the Kolmogorov-Smirnov test. Categorical data are expressed as percentages, and continuous variables as mean $\pm$ standard deviation (SD) or median and interquartile range, as appropriate. A paired 2-tailed Student's $t$-test was used to assess paired data, and an unpaired 2-tailed Student's $t$-test was used to assess unpaired samples. Comparison of more than three normally distributed variables was performed using analysis of variance (ANOVA, with Bonferroni’s post-hoc correction). For the Bland-Altman analysis, paired t-tests versus zero values were used to verify the significance of the biases, and paired t-tests were used to analyse the mean absolute errors of all parameters between healthy subjects and patients. Linear regression was performed to estimate the slope used for correction of myocardial T\textsubscript{1} from blood T\textsubscript{1}. The relationships between corrected and uncorrected mean native myocardial T\textsubscript{1}  were investigated by computing the SD of the mean native myocardial T\textsubscript{1},
and evaluated for difference with a F-test. To investigate whether blood correction improved the discrimination between healthy subjects and patients with CVD's, we calculated the z-scores of the patients in each CVD group with respect to our healthy population for the uncorrected and corrected T\textsubscript{1} maps and compared the average z-scores using a paired 2-tailed Student's $t$-test. Associations between native T\textsubscript{1} values, clinical demographics and LV function were explored by single and multivariate linear regressions. In all cases, p $<$ 0.05 denotes statistical significance. To compute reference values, healthy subjects were used as the study controls and unpaired t-tests were used for comparison.

\section*{Results}
\subsection*{Deep neural network with Bayesian inference}
Table \ref{tab1:dsc_coeffs} reports the Dice scores between automated and manual segmentations evaluated on the cohort of 100 subjects. Overall, the Dice score between manual and automated segmentations was 0.84 for the LV myocardium.\\

The Bland-Altman plot showed strong agreement between the pipeline and manual analysis, see Figure \ref{fig:Fig3_BA_CA_TA}. There was a small negative bias for all of the native T\textsubscript{1} values (-5.04 ms, 5.89 ms and -4.97 for global LV, LVIVS and LVFW respectively). Table \ref{tab2:T1_values} shows the automatically and manually calculated T\textsubscript{1} values within the three regions averaged over the test cohort.  There was no significant difference in mean absolute error between chronic cardiomyopathy patients and healthy volunteers for any of the T\textsubscript{1} values extracted except for LVIVS T\textsubscript{1} values for chronic cardiomyopathy patients. The automatically reconstructed T\textsubscript{1} maps showed a strong correlation with the T\textsubscript{1} values based on manual segmentations (r=0.97) (see Supplementary Figure \ref{fig:Fig3_BA_CA_TA}). \\

Figure \ref{fig:Fig4_ExamplesT1} shows an example of a manual segmentation, the predicted segmentation and the uncertainty map for a healthy subject and for a chronic cardiomyopathy patient. Note that the manual and the automatic segmentations agree well. \\

\subsection*{Quality control}
The balanced accuracy for the QC process was 97.08\%, the sensitivity was 90.08\% and the specificity was 96.44\%.\\

Figure \ref{fig:Fig5_unc_examples} shows examples of uncertainty maps for a cohort of subjects that have been rejected by the QC steps. Note that the QC is able to identify inaccurate data with a wide range of underlying causes such as incorrect planning  (i.e. images (a), (b), (c) and (d) in Figure \ref{fig:Fig5_unc_examples}), motion artefacts (i.e. images (e) and (f) in Figure \ref{fig:Fig5_unc_examples}) or segmentation failure (images (g) and (h) in Figure \ref{fig:Fig5_unc_examples}).

\subsection*{Uncertainty quantification}
To understand the impact of uncertainty in the predicted T\textsubscript{1} values, from the test cohort that contains 50 healthy volunteers and 50 chronic cardiomyopathy patients, we computed the distribution of the global LV T\textsubscript{1} values over the $T$ predicted segmentations. Figure \ref{fig:Fig_AF1_error_t1} shows a graphical representations of the variability of these estimates. The solid lines indicate the mean T\textsubscript{1} values from the test cohort and the shaded region represents one SD of uncertainty.

\subsection*{Myocardial T\textsubscript{1} blood correlation}
The constants from the linear regression model between the myocardial T\textsubscript{1} and the blood measurements were used to correct native myocardial T\textsubscript{1} values according to Equation \ref{eq:blood_correction}. The global R\textsuperscript{2} was 0.28 (which is comparable to that found in \cite{nickander2017blood}) and the mean squared error was 106.4. The linear regression had a slope of -0.35 and an intercept of 935. \\

The mean uncorrected myocardial T\textsubscript{1} in the healthy cohort was 946.44 $\pm$ 61.64, and the mean corrected myocardial T\textsubscript{1} in the healthy cohort was 927.62 $\pm$ 46.41, showing a statistically significant decrease of the SD ((p $<$ 0.001)).

\subsubsection*{Reference values}
From the 14,683 subjects, 156 subjects were rejected by our QC process due to inaccurate T\textsubscript{1} segmentations. Subject characteristics are summarised in Table \ref{tab3:demographics}. Compared to healthy subjects, all patients were in the same age range; patients with obesity had statistically higher body mass index (BMI) and body surface area (BSA); patients with DCM, AF and chronic CAD had significantly lower LVEF (all p $<$ 0.05); patients with HCM, AS and chronic CAD had significantly higher iLVM (all p $<$ 0.05). Native T\textsubscript{1} in three different regions (LV, LVIVS, LVFW) in CVD subjects and in healthy subjects are shown in Table \ref{tab5:ref_vals_add}. Gender did not affect myocardial T\textsubscript{1} values significantly (2-way ANOVA, p = 0.01), and thus only overall T\textsubscript{1} values are reported. \\

Native T\textsubscript{1} values in typical tissue classes in CVD subjects and in healthy subjects for uncorrected and corrected myocardial LVIVS T\textsubscript{1} values are shown in Figure \ref{fig:Fig6_ref_vals_boxplot_uncorrected_corrected_IVS}. Table \ref{tab5:ref_vals_add} shows global and regional native correlated T\textsubscript{1} values. Compared to healthy subjects, patients with DCM, HCM, Takotsubo cardiomyopathy, acute myocarditis and cardiac sarcoidosis had significantly higher native  T\textsubscript{1} values. \\

Finally, Table \ref{tab4:ref_vals_age} shows LVIVS native correlated T\textsubscript{1} values stratified by age by decade (45 to 54, 55 to 64, and 65 to 74 years), and reported as mean and reference ranges (95\% prediction intervals). Note that, in the first age range, i.e. 45-55, there were not enough data to compute the confidence interval for the AS group.

\section*{Discussion}
In this work, we have proposed a fully automated pipeline with a novel quality control step to automatically quantify myocardial tissue from native T\textsubscript{1} mapping, which allows extraction of reference values from large-scale databases. The method is fast and scalable, overcoming limitations associated with current clinical CMR image analysis workflows, which are manual, time-consuming and prone to subjective error. The method has potential to automate T\textsubscript{1} mapping analyses from CMR in clinical practice and research. Using the proposed pipeline we present reference ranges for global and regional myocardial native T\textsubscript{1} in healthy subjects from the UK Biobank dataset and show that blood correction improves discrimination between healthy subjects and patients with CVD.

\subsection*{Automatic analysis with quality control}
We validated our segmentation network by comparing between automated and manual analysis in a cohort of healthy and diseased subjects. Results show a strong agreement for both segmentations (see Table \ref{tab1:dsc_coeffs}) as well as estimated  T\textsubscript{1} values (Figures \ref{fig:Fig3_BA_CA_TA} and \ref{fig:Fig3_BA_CA_TA}). Residual biases in automated T\textsubscript{1} calculations might not necessarily correspond to T\textsubscript{1} estimation errors but might be related to the different methods used to compute T\textsubscript{1} between manual and automatic analysis. Residual biases are within the range of inter- and intraobserver variabilities previously reported \cite{lin2018variability}. Also, residual biases (ranging between -4.97ms and -5.04ms) are consistent between healthy and cardiac patients and are unlikely to have significant clinical impact. The Dice scores we obtained are comparable to previous works \cite{fahmy2019automated}. \\

Quality control techniques are essential to be able to translate deep learning algorithms into a clinical setting. However, many works proposed for the analysis of CMR data have not taken this need into account making it impossible to deploy them for the processing of large-scale databases. In our framework, we employed a novel approach that used the uncertainty information produced by a CNN with Bayesian inference to identify incorrect segmentations, which can be rejected or flagged for revision by an expert cardiologist. We show that this QC process yields over 90\% sensitivity of detecting errors, getting close to clinical standards. One way to further optimize QC in our network would be to incorporate anatomical information into the uncertainty estimation. At the moment, segmentation errors of both RV and LV are accounted for similarly, while the segmentation of the LV is most important for T\textsubscript{1} calculations.

\subsection*{Reference values:}
We used our framework to analyse  native T\textsubscript{1} maps in an unprecedentedly large cohort of healthy volunteers and patients with heart disease. Using this cohort we were able to provide reference values for normal myocardium in aging subjects. T\textsubscript{1} values differ between different scanners, vendors and protocols. Our values can therefore not directly compare to other publications, but they are in agreement with results obtained in smaller cohorts of manual assessment \cite{reiter2013normal, ferreira2014myocardial, piechnik2013normal}. 
We further used our cohort to interrogate T\textsubscript{1} values in patients with 11 different CVDs. We show that for CVDs in which diffuse myocardial disease is prominent (acute myocarditis, cardiac sarcoidosis, HCM, DCM and tako Tsubo disease) we detect significantly larger  T\textsubscript{1} values. For the other CVD's, we show that T\textsubscript{1} values did not significantly change from healthy data. It is likely that the extent of myocardial damage in these groups is less high, explaining the lower T\textsubscript{1} values. There is large variability seen in native T\textsubscript{1} values, which is known to be caused by several factors, including intracardiac blood flow and hematocrite levels. We investigated a previously proposed method for correction of  T\textsubscript{1} values based on blood pool T\textsubscript{1} dynamics \cite{nickander2017blood}. This method has the benefit on working using image data alone, but has not previously been tested in a large cohort of patients. We demonstrate that indeed, discrimination between health and disease improves using blood pool correction. Whether this technique is better than using hematocrite-correction or other methods needs is to be investigated in further studies.
Investigating myocardial disease processes should include additional measures to native T\textsubscript{1}, such as extracellular volume or T\textsubscript{2} imaging. However, the data presented in our study remain valuable. The UK Biobank cohort will contain highly detailed imaging and non-imaging data and follow-up in nearly half a million subjects. On this large scale, new relationships between T\textsubscript{1} and population characteristics can yield important insights into development and progression of CVDs.  

\subsection*{Limitations and future directions:}
A limitation of our work is that the PHiSeg network was trained on a single dataset, the UK Biobank dataset, which is relatively homogeneous. To obtain similar performance in other databases it would be necessary to retrain the networks using a small amount of data. However, the proposed segmentation model and QC steps will remain applicable.\\

Another limitation of this study is the lack of availability of paired LGE and native T\textsubscript{1}-mapping data to assess the correlation between these two measurements, and in which cases T\textsubscript{1}-mapping could provide a better insight into cardiac pathologies. Based on previous studies, it is known that  T\textsubscript{1}-mapping may enable detection of early pathological processes, and serve as a tool for early diagnosis or screening, or differentiation of cardiomyopathies from normal phenotypes. The provided reference ranges could help to identity subjects at risk at an early stage. \\

The T\textsubscript{1} values presented in this study were derived using a single T\textsubscript{1}-mapping technique. It is important to take into account that even within the same T\textsubscript{1}-mapping technique, different versions of sequences can lead to small differences in T\textsubscript{1}-estimations. Therefore, it might not be possible to directly translate the T\textsubscript{1} values derived in this study to other T\textsubscript{1}-mapping techniques. In future work we aim to extend the automatic pipeline to be able to accurately segment T\textsubscript{1}-mapping data from different sequences/vendors, to make the proposed framework generalisable.\\

\section*{Conclusions}
We presented and validated a pipeline for automated quantification of myocardial tissue from native  T\textsubscript{1}-mapping. The proposed method uses the uncertainty of a deep learning segmentation network in a novel QC process to detect inaccurate segmentations. We used the proposed framework to obtain reference values from the largest cohort of subjects to date, which include data from healthy subjects and from patients with the most common myocardial tissue conditions.

\section*{List of abbreviations}
\begin{abbrv}
\item[AF]			Atrial fibrillation 
\item[ANOVA]		Analysis of variance
\item[AS]			Aortic stenosis 
\item[BMI]			Body mass index
\item[BPM]			Beats per minute
\item[BSA]			Body surface area
\item[CAD]			Coronary artery disease
\item[CMR]			Cardiac magnetic resonance
\item[CNN]			Convolutional neural networks
\item[CVAE]         Conditional variational autoencoder
\item[CVD]          Cardiovascular disease
\item[DCM]          Dilated cardiomyopathy
\item[ELBO]         Evidence lower bound
\item[HCM]          Hypertrophic cardiomyopathy
\item[iLVEDV]       Indexed LV end diastolic volume
\item[iLVM]         LV mass index
\item[LGE]			Late gadolinium enhancement
\item[LV]			Left ventricle
\item[LVEF]         LV ejection fraction
\item[LVIVS]		Left ventricle interventricular septum
\item[LVFW]			Left ventricle free-wall
\item[QC]			Quality control
\item[PHiSeg]       Probabilistic Hierarchical Segmentation
\item[ReLU]			Rectified linear unit
\item[ROI]			Regions of interest
\item[RV]			Right ventricle
\item[SAX]			Short axis
\item[SD]			Standard deviation

\end{abbrv}

%%%%%%%%%%%%%%%%%%%%%%%%%%%%%%%%%%%%%%%%%%%%%%
%%                                          %%
%% Backmatter begins here                   %%
%%                                          %%
%%%%%%%%%%%%%%%%%%%%%%%%%%%%%%%%%%%%%%%%%%%%%%

\begin{backmatter}

\section*{Ethics declarations}
\subsection*{Ethics approval and consent to participate}
UK Biobank has approval from the North West Research Ethics Committee (REC reference: 11/NW/0382).
\subsection*{Consent for publication}
Not applicable.
\subsection*{Competing interests}
MS is an employee of HeartFlow, Redwood City, California. All other authors have reported that they have no relationships relevant to the contents of this paper to disclose.

\section*{Acknowledgements}
This research has been conducted mainly using the UK Biobank Resource under Application Number 17806. The authors wish to thank all UK Biobank participants and staff. This research has been conducted using a GPU generously donated by NVIDIA Corporation. 

\subsection*{Funding}
This work was supported by the Wellcome EPSRC Centre for Medical Engineering at Kings College London (WT 203148/Z/16/Z), the EPSRC (EP/R005516/1 and EP/P001009/1) and by the NIHR Cardiovascular MedTech Co-operative. The views expressed are those of the author(s) and not necessarily those of the NHS, the NIHR, EPSRC, or the Department of Health.

\subsection*{Availability of data and materials}
The imaging data and manual annotations were provided by the UK Biobank Resource under Application Number 17806. Researchers can apply to use the UK Biobank data resource for health-related research in the public interest \cite{UKBB}.

\section*{Author's contributions}
Author contribution are as following: conception and study design (EPA, BR, RR and APK); development of algorithms and analysis software (EPA, CFB, EK and MS); data pre-processing (EPA and MS); data analysis (EPA and BR); clinical advice (BR and RR);  manual image annotation (BR); interpretation of data and results (EPA, BR, RR and APK); drafting (EPA, BR and APK);  revising (EPA, BR, CFB, EK, MS, RR and APK). All authors read and approved the final manuscript.

%%  After submission of the .TEX file,                     %%
%%  you will be prompted to submit your .BBL file.         %%

% if your bibliography is in bibtex format, use those commands:
\bibliographystyle{bmc-mathphys} % Style BST file (bmc-mathphys, vancouver, spbasic).
\bibliography{refs}      % Bibliography file (usually '*.bib' )
% for author-year bibliography (bmc-mathphys or spbasic)
% a) write to bib file (bmc-mathphys only)
% @settings{label, options="nameyear"}

% or include bibliography directly:
% \begin{thebibliography}
% \bibitem{b1}
% \end{thebibliography}

%%%%%%%%%%%%%%%%%%%%%%%%%%%%%%%%%%%
%%                               %%
%% Figures                       %%
%%                               %%
%% NB: this is for captions and  %%
%% Titles. All graphics must be  %%
%% submitted separately and NOT  %%
%% included in the Tex document  %%
%%                               %%
%%%%%%%%%%%%%%%%%%%%%%%%%%%%%%%%%%%

%%
%% Do not use \listoffigures as most will included as separate files

\newpage
\section*{Figures}
% Figure 1
\begin{figure}[h!]
\caption{\csentence{Overview of the proposed framework for automatic T\textsubscript{1} map analysis.} A composition of (1) the PHiSeg network for segmenting native T\textsubscript{1} images, (2) a QC for detecting inaccurate segmentation, and (3) T\textsubscript{1} analysis. PHiSeg, Probabilistic Hierarchical Segmentation }
\label{fig:Fig1_pipeline}
\includegraphics[width=\textwidth]{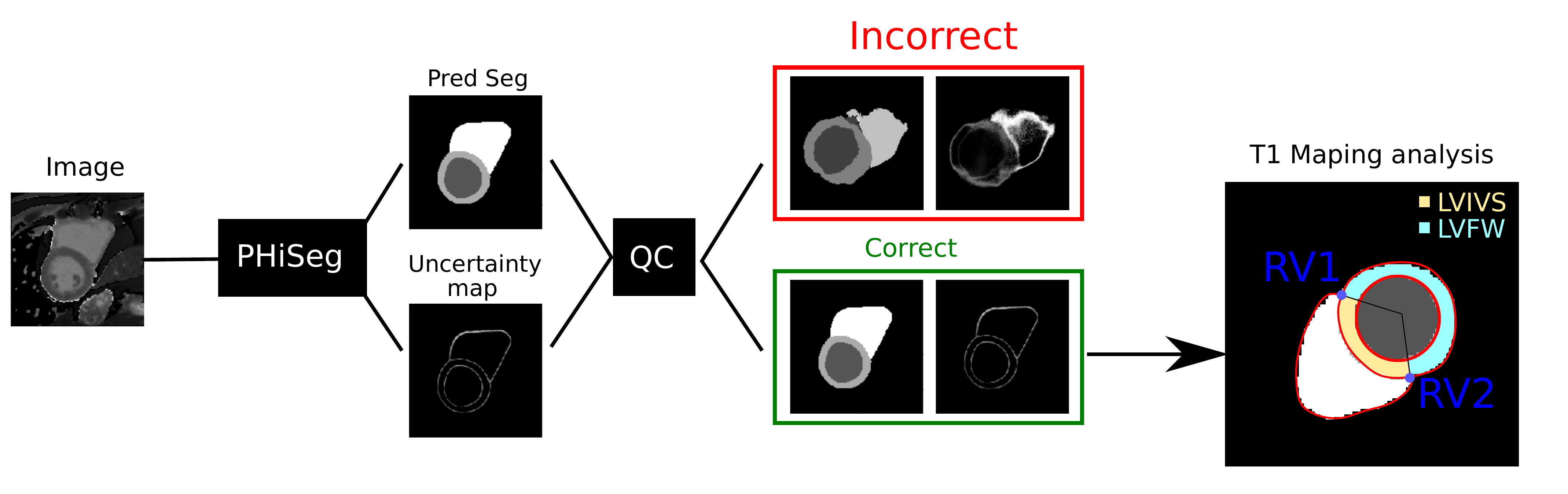}
\end{figure}

% Figure 2
\begin{figure}[h!]
\caption{\csentence{SAX T\textsubscript{1} map with ROIs.} The figure shows regions of interest (ROI) drawn for native myocardial T\textsubscript{1} values of the LVIVS, LV and RV blood pool T\textsubscript{1} values. LVIS, LV interventricular septum.}
\label{fig:Fig2_ROI_T1_corr}
\includegraphics[width=0.5\textwidth]{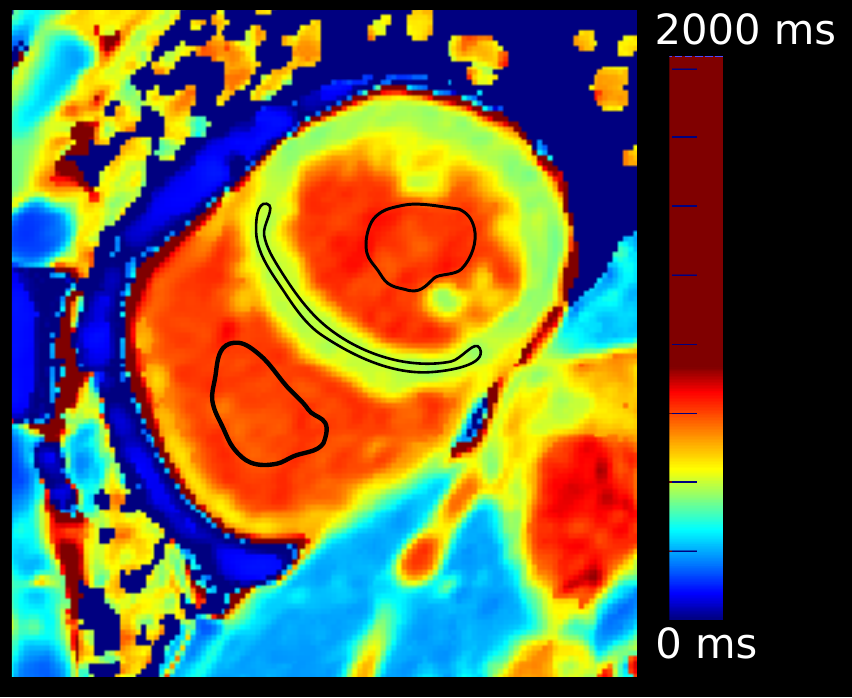}
\end{figure}

% Figure 3
\begin{figure}[h!]
\caption{\csentence{Bland-Altman and correlation plots of the automatic versus manual myocardium T\textsubscript{1} values.} (A, D) global LV native T\textsubscript{1} values, (B, E) LVIVS native T\textsubscript{1} values, (C, F) LVFW native T\textsubscript{1} values. In the Bland-Altman analysis, the grey dotted line represents the mean bias; the red dotted lines the limits of agreement. The p$-$values represent the difference in mean bias from zero using a paired $t$-test. In the correlation plots, the red dotted line represents linear regression line. $r$  is the Pearson’s correlation coefficient. LVIVS, LV interventricular septum; LVFW, LV free-wall.}
\label{fig:Fig3_BA_CA_TA}
\includegraphics[width=\textwidth]{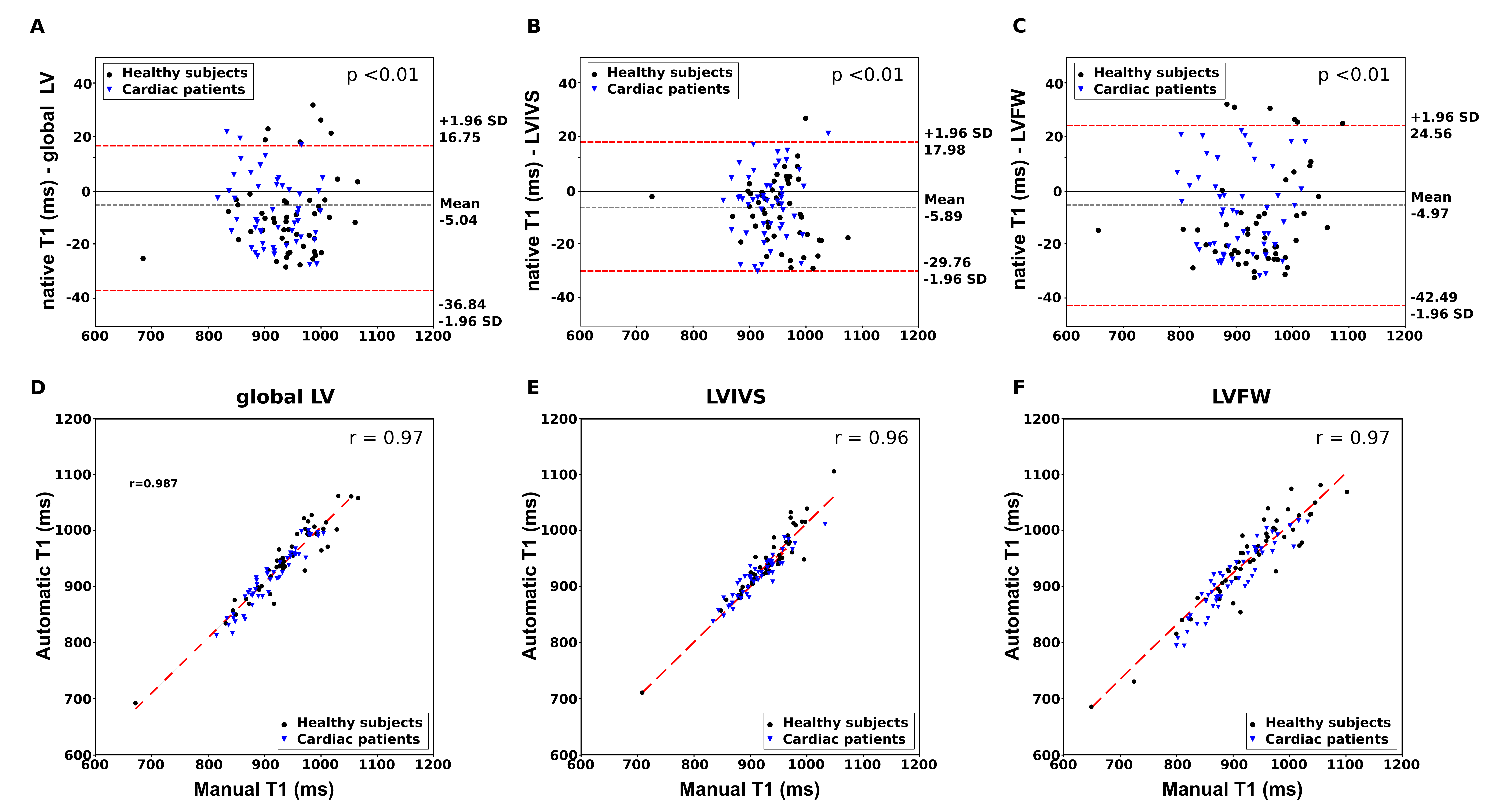}
\end{figure}

% Figure 4
\begin{figure}[h!]
\caption{\csentence{Illustration of the segmentation results for native T\textsubscript{1} images.} The top row shows an example of a healthy subject and the bottom row of a cardiac patient. The cardiac chambers are represented by different colours.}
\label{fig:Fig4_ExamplesT1}
\includegraphics[width=\textwidth]{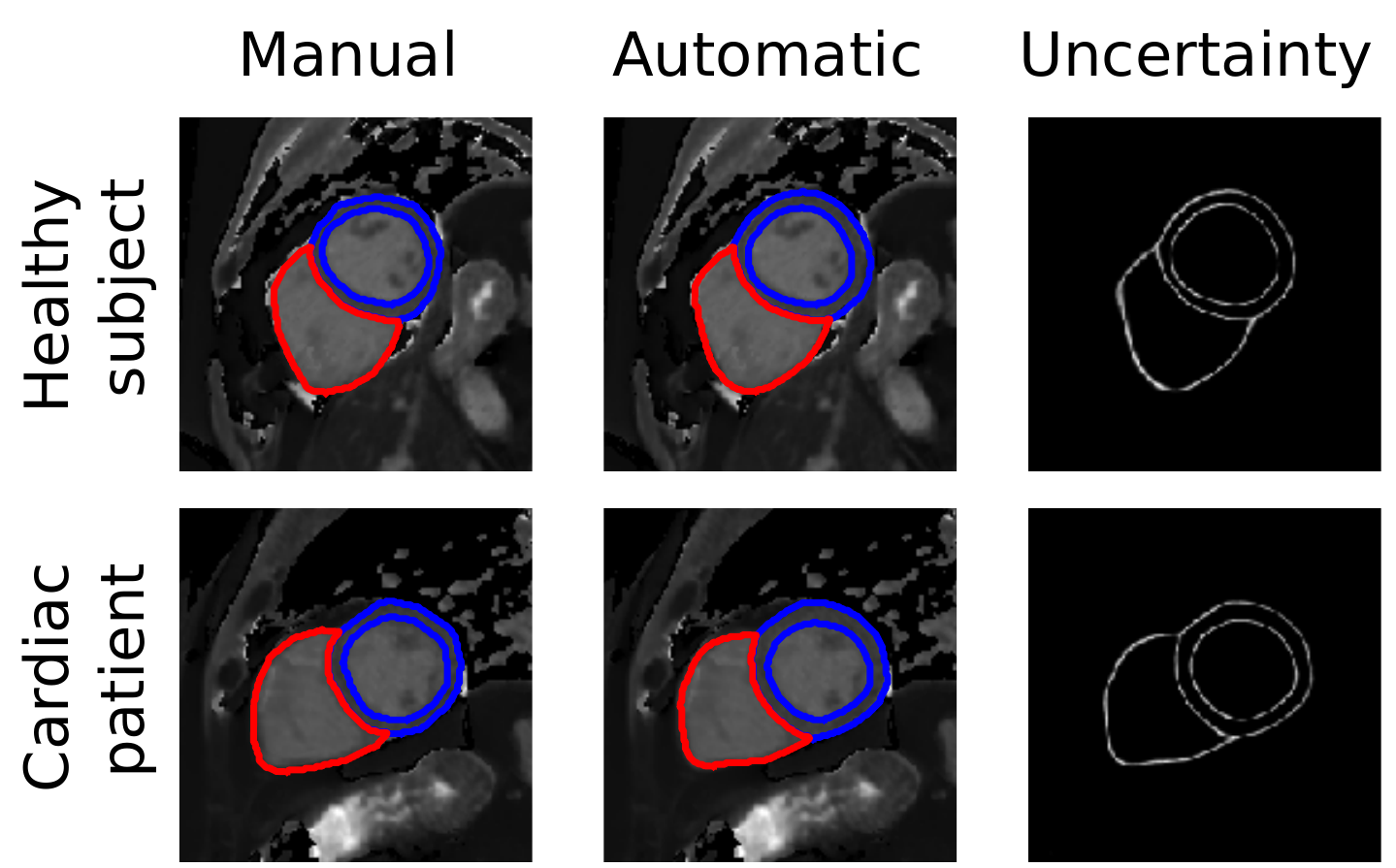}
\end{figure}

% Figure 5
\begin{figure}[h!]
\caption{\csentence{Segmentation and uncertainty map results for selected subjects who were rejected by the QC process}  The left column represents the T\textsubscript{1} mapping images, the middle column the automated segmentation, and the right column the derived uncertainty map. Red arrows indicated the regions with high uncertainty. }
\label{fig:Fig5_unc_examples}
\includegraphics[width=\textwidth]{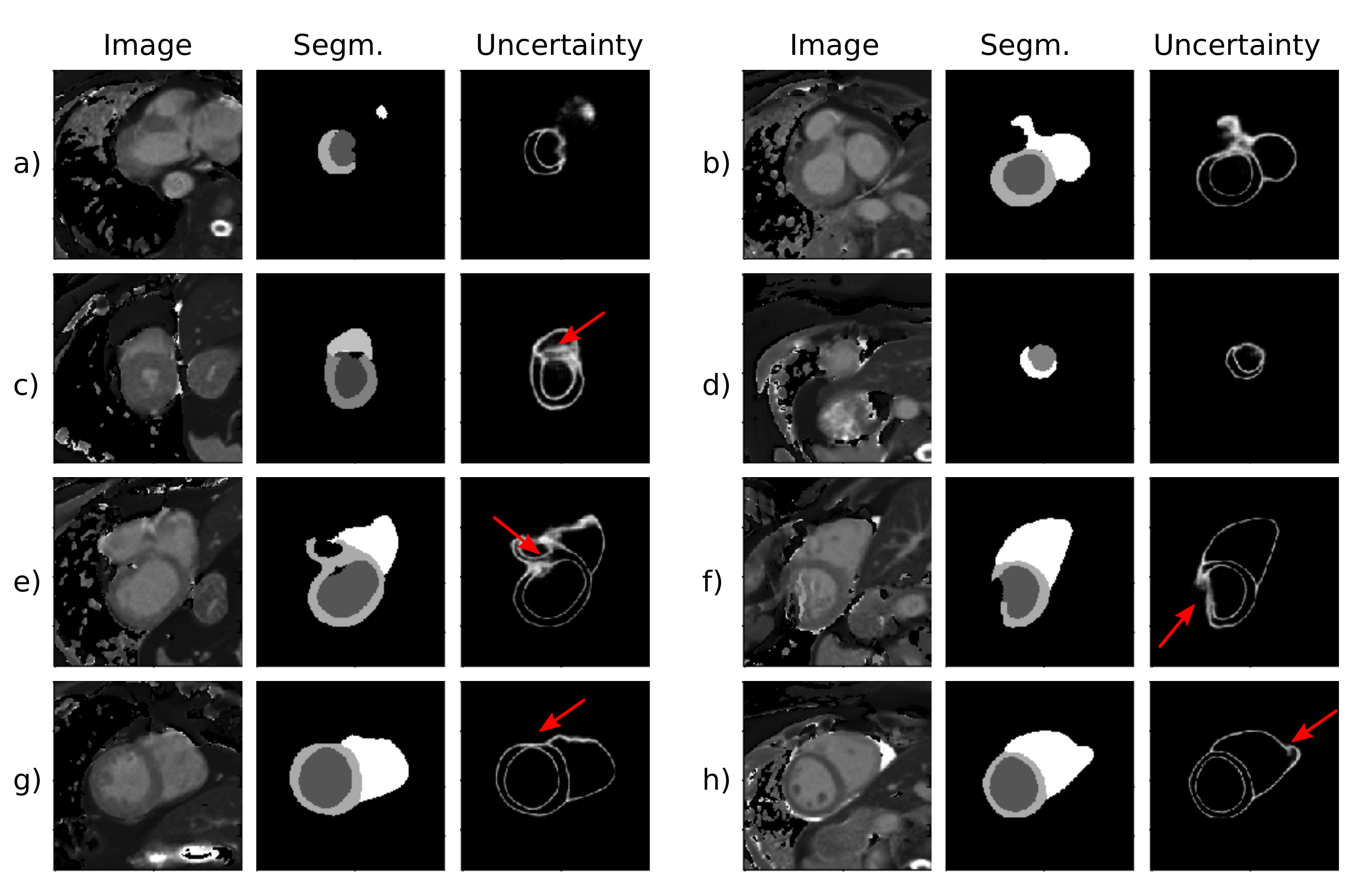}
\end{figure}

% Figure 6
\begin{figure}[h!]
\caption{\csentence{Box plot for uncorrected and corrected LVIVS T\textsubscript{1} for 11 different cardiovascular conditions and healthy subjects.}  Characteristic native LVIVS a) uncorrected and b) corrected T\textsubscript{1} values (1.5 Tesla) for 12 different cardiovascular conditions. Data presented as box and whisker plots with the median, upper and lower quartiles, min and max excluding outliers, and outliers that are more than 3/2 the upper and lower quartiles. Disease names are as per abbreviations list. * denotes values significantly different from healthy subjects and $\dagger$ denotes a significantly decrease of $z$-score from the uncorrected LVIVS T\textsubscript{1} values using a paired $t$-test (all p $<$ 0.05).}
\label{fig:Fig6_ref_vals_boxplot_uncorrected_corrected_IVS}
\includegraphics[width=\textwidth]{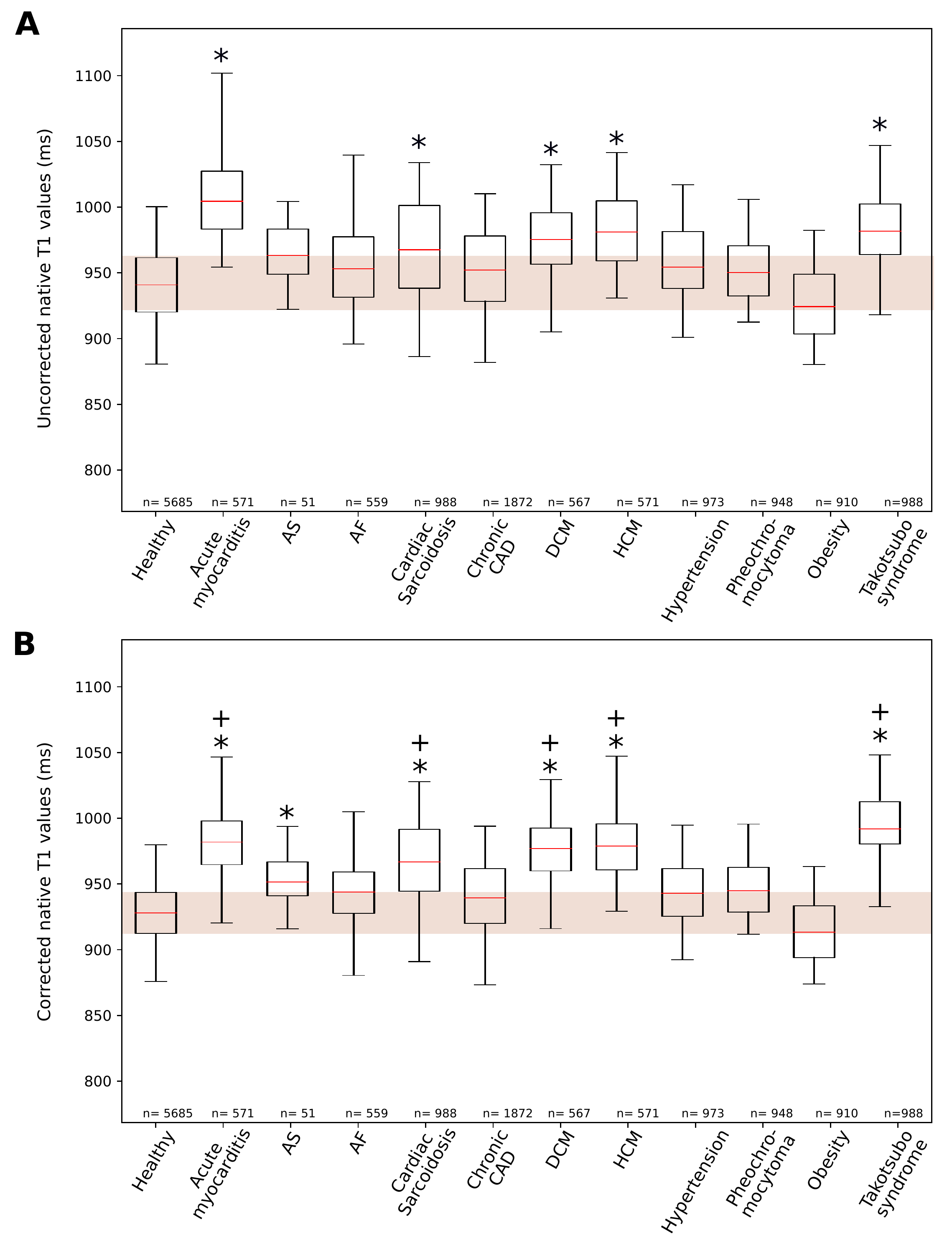}
\end{figure}

%%%%%%%%%%%%%%%%%%%%%%%%%%%%%%%%%%%
%%                               %%
%% Tables                        %%
%%                               %%
%%%%%%%%%%%%%%%%%%%%%%%%%%%%%%%%%%%

%% Use of \listoftables is discouraged.
%%
\newpage
\section*{Tables}
% Table 1
\begin{table}[h!]
\caption{Dice score between automated segmentation and manual segmentation for native T\textsubscript{1} images. Dice values are reported as mean (SD).}
\centering
\begin{tabular}{lccc}
& \multicolumn{3}{c}{\textbf{Mean dice score}}\\ \cline{2-4}
&  LV Blood pool & LV Myocardium & RV blood pool \\ \hline
Healthy  (n=50) & 0.96 (1.81) & 0.83 (3.01) & 0.92 (4.23) \\ 
Chronic cardiomyopathy  (n=50) & 0.95 (1.46) & 0.84 (3.32) & 0.93 (4.06)\\ \hline
Average & 0.95	(1.63) & 0.84 (3.16) & 0.92 (4.14)\\ \hline
\end{tabular}
\label{tab1:dsc_coeffs}
\end{table}

% Table 2
\begin{table}[h!]
\caption{Mean T\textsubscript{1} values for the test cohort. T\textsubscript{1} are reported as mean (SD). Asterisks indicate significant differences between T1 values estimated between manual and predicted segmentation using a paired $t$-test.}
\centering
\resizebox{\textwidth}{!}{
\begin{tabular}{lccc|ccc}
\cline{2-7}
& \multicolumn{3}{c}{\textbf{Healthy}} & \multicolumn{3}{c}{\textbf{Chronic cardiomyopathy}} \\  \cline{2-7}
& global LV & LVIVS & LVFW & LV & LVIVS & LVFW\\ \hline
Manual &  939.33 (65.51) & 947.39 (51.37) & 935.65 (76.59) & 911.18 (47.90) & 930.10 (39.16) & 902.04 (57.81) \\ 
Automatic & 953.99 (66.33) & 959.19 (59.08) & 952.09 (74.16) & 921.40 (51.29) & 935.01 (37.36)* & 912.72 (59.21)\\ \hline
\end{tabular}}
\label{tab2:T1_values}
\end{table}

% Table 3
\begin{table}[h!]
\caption{Baseline characteristics of study subjects included in the analysis for reference values. All values are n (\%) or mean $\pm$ SD. Abbreviations: BMI (body mass index), BSA (body surface area), g (grams), HR (heart rate), kg (kilograms), iLVEDV (indexed LV end-diastolic volume), LVEF (left ventricular ejection fraction), iLVM (indexed LV mass), m (metre). Disease names are as per abbreviations list. * denotes values significantly different from healthy subjects (all p $<$ 0.05).}
\centering
\resizebox{\textwidth}{!}{
\begin{tabular}{lccccccccc}
\hline
& n & Age (years)& Male (n) & BMI (kg/m\textsuperscript{2}) & BSA (m\textsuperscript{2}) & HR (bpm) & iLVEDV(mL/m\textsuperscript{2}) & LVEF (\%) & iLVM (g/m\textsuperscript{2})\\ \hline

Healthy   & 5685  & 60 (28)  & 3056 (53.76) &  25 (4) & 1.84 (0.2) & 61 (9) & 78 (13) & 61 (6) & 45 (8) \\ 

Acute myocarditis & 571 & 68 (6) & 152 (27.19) & 28 (5) & 1.99 (0.2) & 69 (14) & 80 (18) & 57 (13)  & 48 (12) \\

AS & 51 & 68 (7) & 15 (29.41) & 28 (4) & 1.97 (0.2) & 73 (5) &  83 (23) & 59 (6) & 56 (13)*\\

AF & 559 & 68 (6) & 152 (27.19) & 28 (5) & 1.99 (0.2) & 72 (11) & 80 (18) & 51 (9)*  & 51(11)\\ 

Cardiac Sarcoidosis & 988 & 66 (7) & 383 (38.77) & 30 (6) & 2.00 (0.21) & 72 (10) & 77 (18) & 59 (8) & 47 (11)\\

Chronic CAD & 1872 & 67 (7) & 576 (30.77) & 29 (5) & 1.98 (0.2) & 60 (10) & 79 (18) & 48 (9)*  & 55 (11)* \\

DCM & 567 & 68 (7) & 153 (26.98) & 28 (5) & 1.99 (0.2) & 62 (10) & 80 (20) & 39 (9)* & 56(11) \\

HCM & 571 & 68 (6) & 153 (26.80) & 28 (5) & 1.99 (0.2) & 61 (24) & 81 (20) & 57 (9) & 66 (12)* \\

Hypertension & 973  & 63 (7)  & 463 (47.58) &  27 (4) & 1.91 (0.2) & 63 (9) & 77 (13) & 60 (6) & 48 (9) \\

Pheochromocytoma & 948 & 66 (7) & 364 (38.40) & 30 (6) & 2.01 (0.21) & 62 (10) & 78 (18) & 58 (9)  & 46 (15) \\

Obesity & 910 & 66 (7) & 343 (37.69) & 30 (6)* & 2.12 (0.21)* & 62 (10) & 78 (19) & 59 (8)  &  45 (11) \\

Takotsubo cardiomyopathy & 988 & 66 (7) & 383 (38.77) & 30 (6) & 2.01 (0.19) & 58 (10) & 77 (18) & 59 (8) & 44 (13) \\
\hline
\end{tabular}}
\label{tab3:demographics}
\end{table}

\begin{table}[!ht]
\caption{Age-specific native correlated LVIVS ShMOLLI-T\textsubscript{1} reference ranges. Native T\textsubscript{1} reference ranges detailing mean, lower reference limit and upper reference limit by age group. Reference limits are derived by the upper and lower bounds of the 95\% prediction interval for each parameter at each age group.  Abbreviation: LVIVS (left ventricle interventricular septum). Disease names are as per abbreviations list. }
\centering
\resizebox{\textwidth}{!}{
\begin{tabular}{lccccccccccc}
\hline
& \multicolumn{4}{c}{\textbf{45-54}} & \multicolumn{3}{c}{\textbf{55-64}} & \multicolumn{4}{c}{\textbf{65-74}}  \\ \hline
& \textbf{lower} & \textbf{mean} & \textbf{upper} &  & \textbf{lower} & \textbf{mean} & \textbf{upper} &  &\textbf{lower} & \textbf{mean} & \textbf{upper} \\ \hline
Healthy & 793 & 930 & 1066 &  & 793 & 922 & 1052 &  & 781 & 915 & 1049\\

Acute myocarditis & 862 & 973 & 1084 && 876 & 979 & 1082 && 853 & 979 & 1104\\ 
 
AS & & 884 &  && 851 & 957 & 1063 && 840 & 956 & 1072\\ 

AF & 821 & 934 & 1047 && 838 & 941 & 1045 && 816 & 941 & 1066\\ 
 
Cardiac Sarcoidosis & 845 & 950 & 1055 && 857 & 964 & 1071 && 838 & 960 & 1082\\ 

Chronic CAD & 822 & 935 & 1047 && 830 & 940 & 1050 && 814 & 936 & 1059 \\ 
 
DCM & 854 & 965 & 1076 && 868 & 971 & 1074 && 846 & 971 & 1096\\ 

HCM & 854 & 965 & 1076 && 868 & 971 & 1074 && 845 & 971 & 1096\\ 

Hypertension & 797 & 937 & 1077 && 814 & 935 & 1055 && 813 & 941 & 1068\\ 

Pheochromocytoma & 822 & 926 & 1030 && 831 & 937 & 1044 && 813 & 933 & 1054\\ 
 
Obesity & 795 & 900 & 1005 && 807 & 914 & 1020 && 788 & 909 & 1030\\ 
 
Takotsubo cardiomyopathy & 869 & 974 & 1079 && 881 & 988 & 1095 && 862 & 984 & 1106\\ \hline
\end{tabular} }
\label{tab4:ref_vals_age}
\end{table}

% Table 5
\begin{table}[!ht]
\caption{Reference ranges for the native correlated ShMOLLI-T\textsubscript{1} ranges. Reference ranges for the most common myocardial tissue conditions encountered in clinical practice. Abbreviations: LVIVS (left ventricle interventricular septum), LVFW (left ventricle free-wall). Disease names are as per abbreviations list. * denotes values significantly different from healthy subjects (all p $<$ 0.05).}
\centering
\resizebox{\textwidth}{!}{
\begin{tabular}{l c c c c}
\hline
& n & global LV T\textsubscript{1} & T\textsubscript{1} LVIVS & T\textsubscript{1} LVFW\\ \hline
Healthy & 5685  & 927.62 (46.41)  & 922.29 (58.23) &  937.34 (40.75)\\

Acute myocarditis & 571 & 998.39 (50.16)* & 972.28 (59.24)* & 985.59 (41.23)* \\

AS & 51 & 954.49 (35.02) & 943.42 (40.14) & 973.51 (34.08)\\

AF & 559 & 944.28 (49.71) & 935.62 (60.01) & 966.34 (40.13)\\ 

Cardiac Sarcoidosis  & 988 & 972.33 (48.35)* & 957.07 (57.40)* & 984.18 (40.46)*\\

Chronic CAD & 1872 & 942.46 (47.18) & 933.46 (56.58) & 959.25 (38.61)\\

DCM & 567 & 974.72 (47.18)* & 968.71 (59.11)* & 990.38 (39.16)* \\

HCM & 571 & 989.41 (50.16)* & 968.68 (55.55)* & 989.61 (41.29)* \\

Hypertension & 973  & 943.75 (49.21)  & 935.68 (58.24) &  958.82 (36.75) \\

Pheochromocytoma & 948 & 944.23 (48.47) & 935.94 (56.68) & 963.02 (41.60)\\

Obesity & 910 & 916.35 (47.73) & 907.78 (56.94) & 932.45 (41.71) \\

Takotsubo cardiomyopathy & 988 & 991.82 (48.61)* & 983.10 (57.45)* & 1009.81 (41.68)* \\

\hline
\end{tabular}}
\label{tab5:ref_vals_add}
\end{table}

%%%%%%%%%%%%%%%%%%%%%%%%%%%%%%%%%%%
%%                               %%
%% Additional Files              %%
%%                               %%
%%%%%%%%%%%%%%%%%%%%%%%%%%%%%%%%%%%
\newpage
\section*{Additional Files}
\begin{figure}[h!]
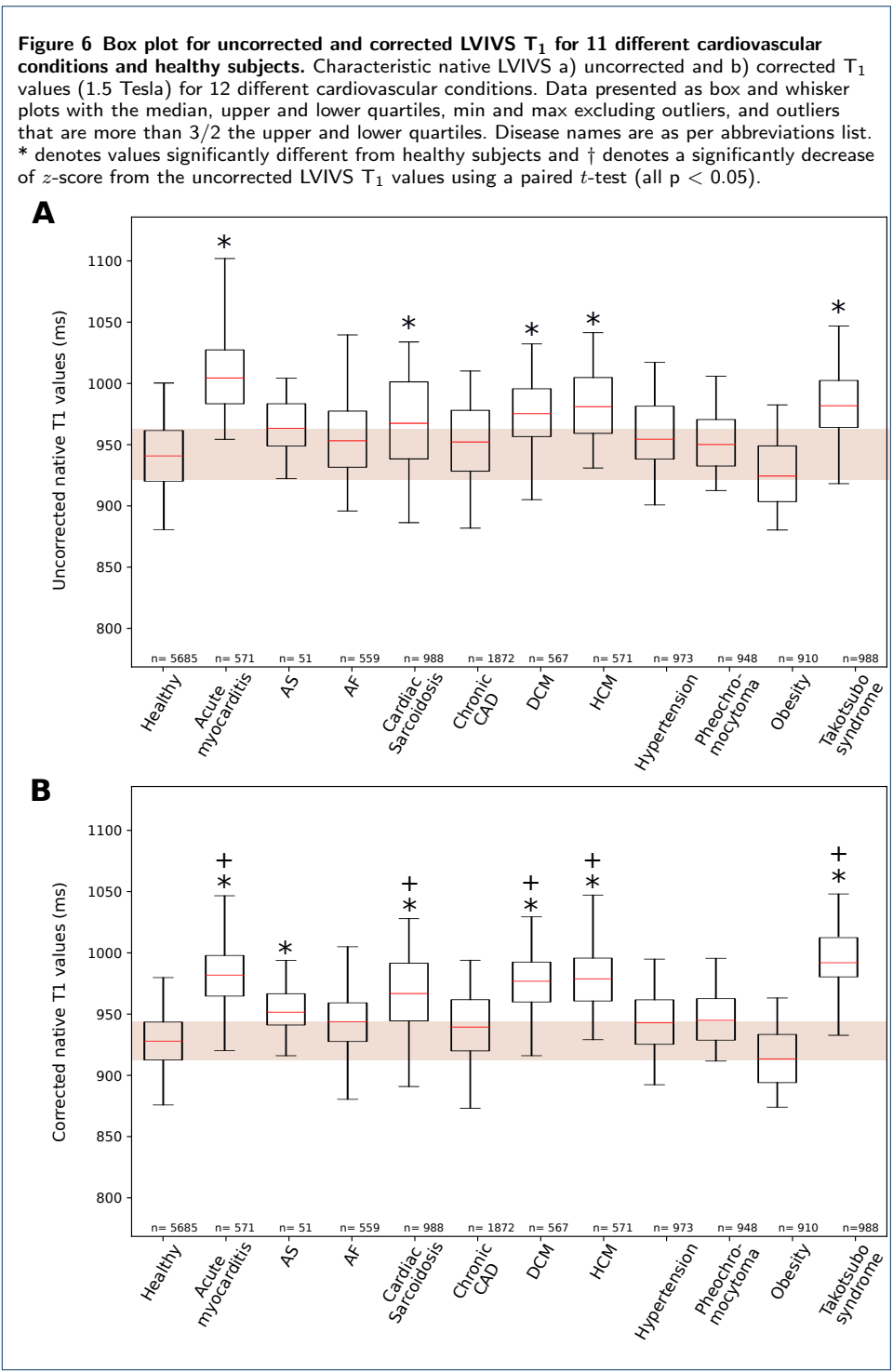

\caption{\csentence{Uncertainty plot for predicted global global LV T\textsubscript{1}.}  Data presented in ascending order of predicted T\textsubscript{1} showing mean T\textsubscript{1} value and SD of uncertainty for each subject. Colours represent the different groups in the test cohort. }
\label{fig:Fig_AF1_error_t1}
\end{figure}

\end{backmatter}
\end{document}